\documentclass{aa}  

\usepackage{graphicx}
\usepackage{txfonts}
\usepackage{xcolor}
\usepackage{dirtytalk}

\usepackage[normalem]{ulem}
\usepackage[colorlinks=true,
            citecolor=blue,
            linkcolor=blue,
            urlcolor=blue]{hyperref}
\newcommand{\kms}{km s$^{-1}$}

\newcommand{\moh}{[\mathrm{M/H}]}
\newcommand{\aom}{[\mathrm{\alpha/M}]}

\begin{document}

   \title{The chemical side of the Milky Way warp.} 

   \subtitle{The $[\alpha/\text{M}]$ abundance from Red Giant Branch stars in Gaia DR3 reveals the warped distribution of the disc.}

   \author{Mauro Cabrera-Gadea
          \inst{1,2}
          \and
          Danny Horta\inst{3}  
          \and 
          Pau Ramos \inst{4}
          \and
          Cecilia Mateu\inst{1,2}
          \and
          Adrian M. Price-Whelan\inst{5}
          \and
          Emma Tasso\inst{6}
          }

   \institute{$^1$ Departamento de Astronom\'{i}a, Instituto de F\'{i}sica, Universidad de la Rep\'{u}blica, Igu\'a 4225, CP 11400 Montevideo, Uruguay\\
   $^2$ Programa de Desarrollo de las Ciencias Físicas (PEDECIBA), Isidoro de María 1614, Montevideo, Uruguay\\
   $^3$ Institute for Astronomy, University of Edinburgh, Royal Observatory, Blackford Hill, Edinburgh, EH9 3HJ, UK\\
   $^4$ National Astronomical Observatory of Japan, Mitaka-shi, Tokyo 181-8588, Japan\\
   $^5$ Center for Computational Astrophysics, Flatiron Institute, 162 Fifth Avenue, New York, NY 10010, USA\\
   $^6$ The Graduate Center, City University of New York, 365 Fifth Avenue, New York, NY 10016, USA\\
              \email{mauro.cabrera@pedeciba.edu.uy}
         }

   \date{Received ...; accepted ...}
 
   \abstract
   {The outskirts of the Milky Way disc are warped. Previous works mainly characterised this warp by analysing the distribution function, $f$, that is prone to selection-function biases because it models the density of stars in phase-space. However, large-scale spectroscopic surveys now also supply element abundances for vast samples of stars; with these data, various methods have been able to trace features in $f$ by means of tracing the chemical abundance gradient distribution in phase space, which is less sensitive to spatial selection function effects. This new approach has yet to be exploited to characterise the warp.}
   {Our goal is to empirically trace the Galactic disc's warp by means of the gradient of alpha-to-metallicity, $\aom$, element abundances in the plane of vertical kinematics.}
   {We use an existing catalogue of red giant branch stars with $\aom$ abundances from \textsl{Gaia} XP spectra, combined with \textsl{Gaia} DR3 astrometry and radial velocities. We use the Orbital Torus Mapper method to trace the distribution of $\aom$ elements in the $L_z,\phi,z,v_z$ space; this allows us to trace the centre of the distribution in the $z,v_z$ plane for different cells of $L_z,\phi$ across the disc.}
   {For the first time, the Milky Way warp is empirically determined using element abundance gradients. The structured chemical distribution in the $z,v_z$ plane is found to be preserved in the outer disc, mapping $z$ and $v_z$ up to $L_z\approx3300\text{ kpc km/s}$ ($R_g\approx14\text{ kpc}$) for a wide range in azimuth, in the anticentre direction. The centre of the $\aom$ distribution in $z$ closely follows the observed mean $z$ for Classical Cepheids, with a similarly twisted line of nodes. In the case of $v_z$, the observed signal shows some ``second order'' differences with respect to the Cepheids. A characterisation of the distribution of $\aom$ in the $z,v_z$ plane at the warped region shows some degree of lopsidedness which can be associated with an early stage of phase mixing. Finally, we show that the mean $v_z$ as a function of $L_z$ for mono-$\aom$ populations decreases with increasing $\aom$ and becomes more monotonic for $\aom$-rich populations (kinematically hotter), while $\aom$-poor (colder) populations show a signal of shorter wavelengths in $L_z$.}
   {Our results indicate that the relation between the distribution function and element abundance gradients can be traced with current data even in the warped region. They also indicate that the disc is highly structured, allowing the exploration of the warp not only as the mean $z$ and mean $v_z$, but also the distribution on the $z,v_z$ plane. Moving forward, any model attempting to explain the origin of the warp, which still eludes us, has a new benchmark to compare against thanks to the new approach used in this work for characterising its morphology, kinematics and chemistry.} 
   
   \keywords{Galaxy: disk --
                Galaxy: structure --
                Galaxy: kinematics and dynamics
               }

   \maketitle

\section{Introduction}

The Milky Way galaxy has its gaseous and stellar disc warped \cite{} as many other disc galaxies \citep{Sanchez-Saavedra_1990_Statistical_Study_of_Optical_Galactic_WARPS}. This undulation makes the bulk of stars go up and down in a coordinated fashion as they orbit the Galaxy. The warp in our Galaxy has been known since the late 1950s \citep{Burke_HI_Warp_1957}, but its dynamical origin is still puzzling the community. Some channels proposed include: gas accretion \citep{Roskar_Acretion_warp_2010}, the dark matter distribution \citep{Sparke_Making_WARPS_from_bars_haloes,Ideta_warps_in_prolate_haloes,Jeon_2009_Warps_Triaxial_halos,han_halo_warp_2023}, or interaction with external galaxies like the LMC or the Sagittarius dwarf spheroidal galaxy \citep{Weinberg_Magellanic_Origin_Warp_2006,Chequers2018,Semczuk_Tidally_induced_warps,Bland-Hawthorn_2020_Galactic_seismology,Poggio_Warp_evolving,Stelea_RowdyNeighbors_2024}.

To better constrain any dynamical origin of the warp it is essential to characterise its structure and kinematics. The stellar warp has been studied by various methods, such as star counts \citep{Evolution_disc_shape_Amores_2017}, modelling direct measurements of the mean vertical height \citep{Warp_Pulsars_Yusifov_2004, Skowron2019,Cheng_2020_Warp_patern,Cabrera_Gadea_etal_2024,Cabrera_RRLyrae_2025}, or via inference of the vertical velocity profile \citep{Warp_RedClump_Lopez_Corredoira_2002,Cheng_2020_Warp_patern,Wang_LAMOST_Disk_Warp_RC_2020,McMillan_Wap_2024}. Although using different approaches, these methods have mainly sought to recover moments of the distribution function $f$, such as the mean vertical height, $z$, or the mean vertical velocity, $v_z$, weighted by the reconstructed $f$. Naturally, when working with $f$, one must always take into account the selection function, particularly when modelling the mean vertical height, where it introduces the largest biases \citep[see Appendix B in][]{Cabrera_Gadea_etal_2024}. 

Thanks to surveys such as \textsl{Gaia}, \textsl{LAMOST}, and \textsl{APOGEE}, with their great quality and amount of data mapping the phase space of our Galaxy, a new approach for analysing properties of $f$ has been put into play without the need to reconstruct $f$ itself. 
By making use of invariant ``stellar labels'' such as ages or element abundances, various works have been able to explore the dynamics of the Milky Way disc, thanks that various features in $f$ are present in the distribution of the stellar labels in phase space. The most notorious example is the phase spiral over the plane $z,v_z$ \citep{Antoja_2018_Phase_spiral}, a clear signal of phase mixing in the disc after a disequilibrium event. This perturbation in $f$ has been revealed in various stellar labels as the mean stellar age, the mean iron abundance, and [Mg/Fe] abundance \citep{Hawthorn_PS_dissected_GALAH_2019,Frankel_Iron_Snail_2025,Alinder_2023_rotation_phase_spiral}. As shown by \citet[][Section 2.3]{Frankel_Iron_Snail_2025} this link between $f$ and some stellar labels is a consequence of the collisionless nature of galactic dynamics. The key advantage of this link is that stellar labels are less prone to bias than $f$ by the spatial selection function of the survey, as demonstrated also by \cite{Adrian_OTI_2025}. This makes stellar labels very useful unbiased tracers of the kinematics, providing a key advantage over modelling the density of stars with $f$ itself. 

By means of this mapping between $f$ and some stellar labels, previous works determined vertical acceleration and surface mass density for a wide region of the disc thanks to their ``Orbital Torus Imaging'' method (OTI: \citealp{Adrian_OTI_2021,Adrian_OTI_2025,Danny_OTI_az_2024}). These works  exploited the fact that in a steady-state and axisymmetric system, the isocontours of $f$ in the $z,v_z$ plane for a volume of the disc are well mapped by isocontours of mean element abundance (namely, $\langle\text{[$\alpha$/Fe]}\rangle$). This highly structured chemo-dynamical relation of the Milky Way disc allows them to reconstruct the orbits in the vertical direction and therefore recover the local gravitational potential. The validity of the OTI method when the Galaxy experiences a certain amount of disequilibrium was explored in \citet{OTI_on_FIRE_2025_Oeur}, by applying OTI in a warped disc galaxy from the \texttt{FIRE} suite of cosmological hydrodynamical simulations; the results from this study showed that OTI is able to recover the vertical acceleration profile even in a dynamically disturbed system. More importantly for this work, \citet{OTI_on_FIRE_2025_Oeur} showed that OTI can also map the warp signal locally. These results remark how even in a Galaxy not in (dynamical) equilibrium, like the Milky Way, the structured chemical-dynamical relation of the disc can still be traced with stellar labels.

The chemistry and the warp have been linked by \cite{Sun_2025_Fe_warp} empirically, showing that the midplane of the iron abundance from red clump stars traces a warped disc with a similar amplitude to the Classical Cepheids, although a detailed view of features like the twisted line of nodes and the kinematics of the warp remains to be unveiled. In this work, we set out to use the OTI method for exploring the Milky Way warp by means of analysing the element abundance gradient across the plane of vertical kinematics using red giant branch stars from \textsl{Gaia} DR3 \citep{GaiaDR3}.

This paper is organized as follows: In Section~\ref{s:sample} we present the data used and the quality cuts used. In Section~\ref{s:method} we explain the OTI method and its application to our data. In Section~\ref{s:results} we show the relation between the element abundances and the warp, and describe how OTI works to trace it. In Section~\ref{s:discussion} we discuss our findings, before presenting our conclusions and discussing future work in Section~\ref{s:conclusion}.

\begin{figure*}[t]
	\includegraphics[width=\textwidth]{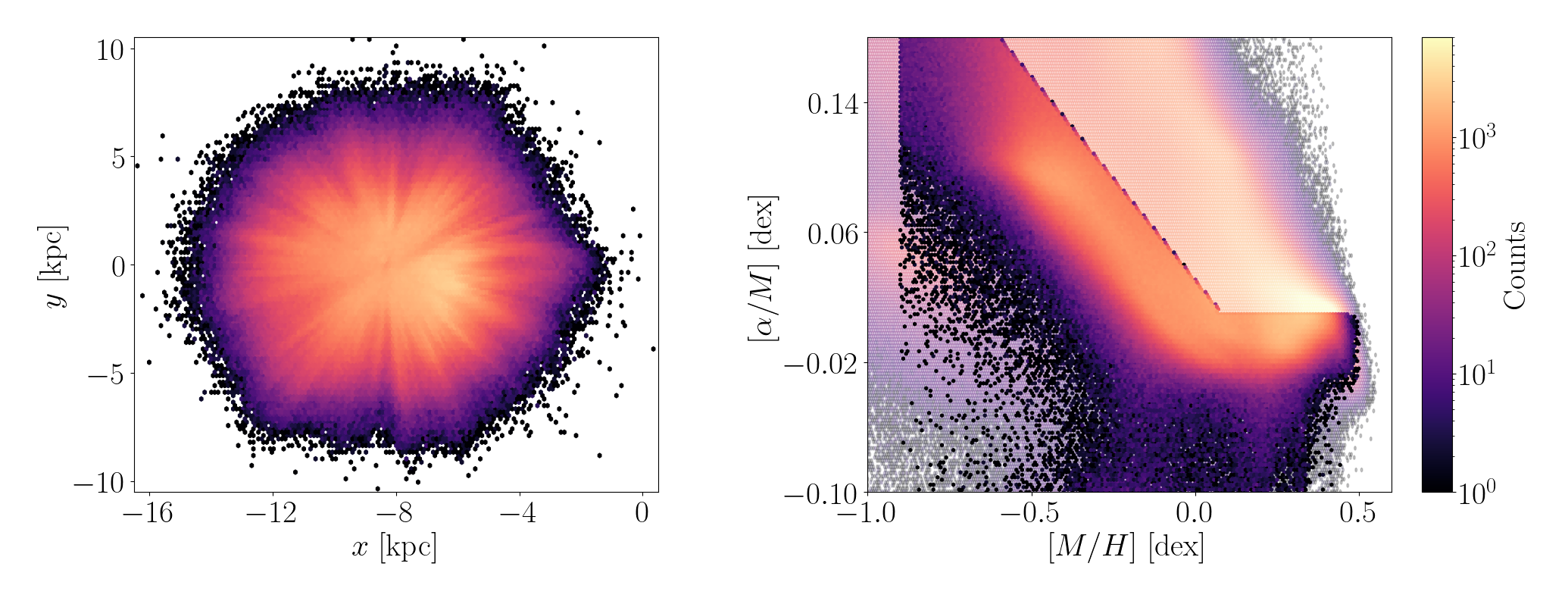}
    \caption {Summary plots of our red giant branch star sample. The left panel shows a 2D histogram of the ``face-down'' view of the disc plane. The right panel shows the Tinsley-Wallerstein diagram, (i.e., $[\text{M}/\text{H}] -[\alpha/\text{M}]$ plane), and the cut we employ to select our low-$\alpha$ disc sample. The full sample is shown as the faint background histogram.} 
    \label{fig:sample} 
\end{figure*}

\section{Sample}\label{s:sample}

\noindent The \textsl{Gaia} mission/survey \citep{The_Gaia_mission_2016} has so far delivered detailed measurements of sky positions, proper motions, and parallaxes for $\approx1.5$ billion stars, limited only by their apparent magnitudes (\textsl{Gaia} $G<20.7$). For this work, we used data from the third \textsl{Gaia} data release \citep[DR3:][]{GaiaDR3}, supplemented by stellar parameter and element abundances (namely, $T_{\mathrm{eff}}$, $\log~g$, [M/H], and [$\alpha$/M]) determined by \citet{Li_AspGap_RGB_2024} via the \texttt{ASPGAP} model using the low-resolution $BP-RP$ (or XP) spectra \citep{DeAngeli_GaiaXP_2023,Montegriffo_GaiaXP_2023}.  \citet{Li_AspGap_RGB_2024} determined stellar parameters and element abundances for \textsl{Gaia} stars by training a model using an overlapping sample of \textsl{Gaia} red giant branch stars with optical low-resolution XP spectra as well as \textsl{APOGEE} near-infrared spectra, stellar parameters, and element abundances, and then applying this model to the \textsl{Gaia} data. In total, this sample of \textit{Gaia} red giant branch stars with full six-dimensional phase-space measurements, stellar parameters, and [M/H] and [$\alpha$/M] element abundances is comprised of $\approx23$ million stars across the Milky Way.

We convert the 6D phase-space observables into a Galactocentric cylindrical coordinates using the \texttt{Astropy}\citep{astropy2018} package, assuming a distance from the Sun to the Galactic centre of $8.27$~kpc \citep{GRAVITY_collab_2022_Rsun}, $z_{\odot}=0.027$~kpc \citep{ZsolarChen}\footnote{We have assessed how including/removing this solar vertical assumption affects our results, finding that it has no effect.} and a solar velocity of $v_\odot = (11.10, 248.5, 7.25)$ km s$^{-1}$ \citep{V_sun_Schonrich_2010,Reid_2020}. In this frame, negative $v_\phi$ corresponds to prograde disc rotation.

Our parent sample is comprised of stars that satisfy the following quality cuts:
\begin{description}
    \item[\textbf{High quality astrometry:}] photo-geometric distances \citep{Bailer-JonesGDR3} with $\varpi/\sigma_\varpi>5$ and $\text{RUWE}<1.4$, as recommended by \cite{Lindegren_2021_GEDR_Astrometric_Solution}.
    \item[\textbf{Red giant branch stars:}] Although the sample by \cite{Li_AspGap_RGB_2024} is restricted to giants, we further restrict our sample to $1.5<\log g<3.5$, as in \cite{Danny_OTI_az_2024}, limited to a narrow range in surface gravity, and thus mitigate as much as possible any trends of element abundances with distance/brightness.
    \item[\textbf{Low-$\alpha$ disc stars:}] \texttt{ASPGAP} element abundances that fall on the low-$\alpha$ sequence in the $\aom$--$\moh$ plane (Figure~\ref{fig:sample}), which satisfy the following relation:
    \begin{equation*}
    \begin{cases}
    \aom <0.18, & -0.9<\moh < -0.6, \\[2pt]
    \aom < - 0.25\,\moh + 0.03, & -0.6 \le \moh < 0.07, \\[2pt]
    \aom <0.01, & \moh \ge 0.07.
    \end{cases}
    \end{equation*}
    \item[\textbf{Reliable element abundance measurements:}] 
    We also restrict the sample to $\aom>-0.1$ since stars below this value have unrealistic abundances for the disc and are affected by extinction. Finally, we also drop those stars with unrealistically low uncertainties in $\aom$ abundance by taking $\sigma_{\aom}>0.005~[\text{dex}]$
    \item[\textbf{Stars with thin disk kinematics:}] Vertical velocities within $|v_z|<90~\text{km/s}$ and vertical positions within $|z|<3$~kpc. We also selected stars with low radial velocities $|v_R|<50~\text{km/s}$.
    
\end{description}

In Figure~\ref{fig:sample} we illustrate the properties of our final sample by showing a 2D histogram of the distribution of stars over the disc plane (left panel) and $[\text{M}/\text{H}],[\alpha/\text{M}]$ plane (right panel).

\section{Method}
\label{s:method}

To analyse the distribution of chemical abundances in the plane $z,v_z$ in different volumes of the disc, we make use of the OTI \citep{Adrian_OTI_2021} framework. As noted in \citet{Danny_OTI_az_2024}, \citet{Adrian_OTI_2025}, and \citet{Danny_OTI_aR_2026}, the OTI framework's main idea is that, for a galaxy in (approximate) steady-state, the distribution of intrinsic stellar invariants (for example, element abundances of stars) for a tracer population can only depend on dynamical invariants and not on time or orbital phase. For a small region of the Milky Way disc and a 2D slice of phase space (e.g., vertical kinematics $\boldsymbol{w} = (z, v_z)$), OTI assumes that contours of constant mean element abundance should correspond to orbital trajectories of constant vertical action, $J_z$, and that the gradient in mean element abundance as a function of phase space coordinates correlates with a foliation of orbits with different $J_z$ in the 2D kinematic plane. Thus, if each orbit is populated by a grouping of stars with an approximately identical value of a stellar invariant, then we can reconstruct the isocontours of the distribution function, $f$, by just mapping the isocontour of the mean element abundances in phase space. For more details on the analytical description of the OTI model and the assumptions this framework requires, we refer the reader to \citet{Adrian_OTI_2025} and \citet{Danny_OTI_az_2024}. 


Recent works testing the validity of the OTI method and its assumptions under different levels of disequilibrium have been performed using either numerical simulations \citep{Adrian_OTI_2025} or cosmological hydrodynamical simulations \citep[\texttt{FIRE}:][]{OTI_on_FIRE_2025_Oeur}. These results attest to the reliability of the OTI method to effectively measure key dynamical properties of the disc, such as vertical acceleration (i.e., the force law) or enclosed mass density. In addition, when modelling the plane of vertical kinematics $(z-v_z)$, the OTI framework is also capable of measuring two additional important quantities that can help describe the kinematic structure of the Galactic disc: the midplane position, $z_0$, and the midplane vertical velocity, $v_{z0}$. \citet{Danny_OTI_az_2024} used this principle to measure the Sun's vertical position and velocity. However, in practice, with enough data, this method can be applied to measure the (orbital) midplane position and vertical velocity across the Milky Way disc; this provides an avenue to empirically map the Galactic disc's warp and vertical velocity structure. 

\begin{figure*}
	\includegraphics[width=\textwidth]{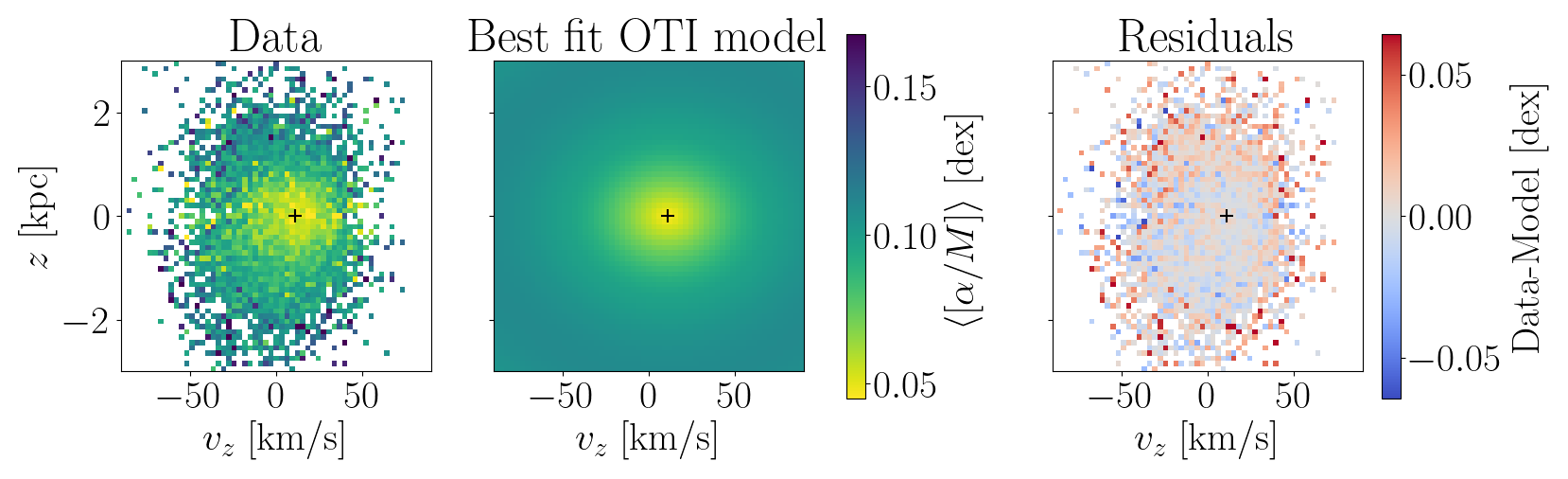}
    \caption{ $z$ versus $v_z$ plane for a particular angular momentum ($L_z=2900\text{ kpc km/s}$) and azimuth ($\phi=180^\circ$) bin. \textit{Left}: Distribution of the mean $\aom$ (colour scale) for the data.  \textit{Middle}: best fit OTI model. \textit{Right}: Distribution of model fit residuals  (data $-$ model). The black cross shows the centre of the abundance distribution of the OTI fit (i.e., ($z_0,v_{z0}$)). }
    \label{fig:OTImethod}
\end{figure*}

In this work, we build on previous OTI results and set out to map the $(z-v_z)$ plane across the Milky Way disc with the aim of measuring the chemical midplane position, $z_0$, and the mean vertical velocity at the midplane, $v_{z0}$ in small bins of orbital radius and azimuth. As we are not interested in measuring the vertical acceleration profile, our method requires fewer assumptions. More specifically, the orbital centre ($z_0, v_{z0}$) is determined by the location about which the label contours are most nearly symmetric, or where the inflexion point in the element abundance gradient occurs w.r.t. $z$ or $v_z$. Thus, it is fixed by the geometry of the abundance distribution in the $(z-v_z)$ plane, rather than by the requirement that those contours be genuine orbital tori, needed to foliate the orbital plane to determine the vertical acceleration. In theory, we could have instead just set out to measure the saddle point of the element abundance distribution (similar to the method in Tasso et el in prep). However, since we have the OTI method at hand and this has been extensively tested, we chose to use this framework to perform these measurements.

To perform our measurements, we divide the disc into bins of angular momentum, $L_z$, $110~\text{kpc km/s}$ wide (corresponding roughly to $0.5$~kpc in radius). We also divide our sample into three Galactocentric azimuthal bins with means of $160,180,200\deg$ and $30\text{~deg}$ wide. We have found that this configuration provides the best trade-off between spatial resolution and number of stars in each bin, ensuring our measurements are robust. An example of the data (left), the model fit (middle), and the residuals (the difference between the model and the data, right) is shown in Figure~\ref{fig:OTImethod}. In this Figure, each panel shows the pixelated data for a randomly selected angular momentum and azimuth bin in the vertical kinematics plane $(z-v_z)$, coloured by the average value of $\aom$. As can be seen from Figure~\ref{fig:OTImethod}, the OTI model is able to capture well the overarching smooth element abundance gradient distribution, despite some underlying disequilibrium features (on the order of $<10\%$). Importantly for this work, our method is able to pin point well the orbital centre ($z_0,v_{z0}$), highlighted in Figure~\ref{fig:OTImethod} as a black cross. Lastly, we choose to use $\aom$ instead of $\moh$ since $\aom$ has been shown to strongly correlate with vertical kinematics in the Milky Way disc \citep{Danny_OTI_az_2024,Danny_OTI_aR_2026} due to its ordered chemical-dynamical structure \citep[e.g.,][]{Ness2019}.

Before closing this Section, it behoves us to mention the effects of selection function on our results. \cite{Adrian_OTI_2025} showed how a reasonable selection function over a mock catalogue does not affect OTI's results, since the spatial selection function does not strongly affect the gradient of mean element abundances so long as enough stars have been measured and the gradient can be recovered. For our case, the only possible selection function effect that could bias our method would be caused by stars at different distances, and thus different brightness or $\log~g$, having a correlation with their element abundance ratio. We have ensured that the element abundance gradients we model in the plane of vertical kinematics still stand when using a volume limited sample, finding that our results remain unchanged. Thus, given this test, and the fact that we limit our sample to a reasonably limited brightness range (namely, $ 1.5< \log~g <3.5 $), we are confident that this effect has been appropriately mitigated and that our results are not biased by element abundance trends with distance or brightness. See Appendix \ref{Ap:A} for further details. 

We also explored whether differential extinction along the $z$ direction could affect the $\aom$ values in a way that might bias the OTI fits. For this, we took a disc cell with asymmetric extinction relative to the plane but in a non-warped region (so we know that $z_0$ should be $\approx0$~kpc); this way, we further divide into the sample with higher extinction $A_V$ than the median and the sample with lower $A_V$ than the median (the extinctions are taken from \cite{Green2019}). Finally, we fit OTI to both samples, and both differ by $10$~pc in their $z_0$, which is lower than the changes by bootstrap realisation from the full sample, so we do not consider that differential extinction drives a major bias in the OTI measurements.

\begin{figure*}
	\includegraphics[width=\textwidth]{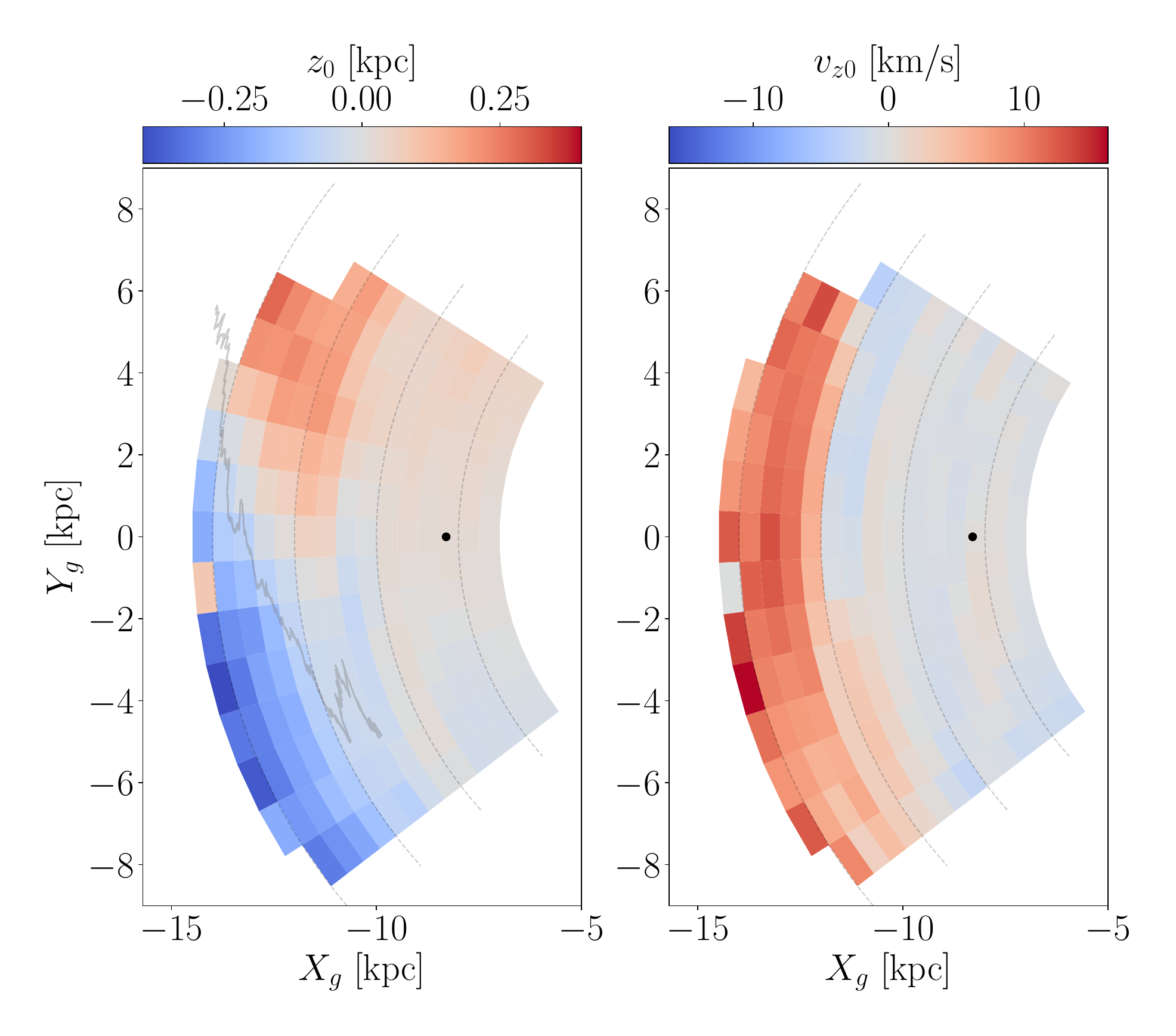}
    \caption{The colour scale shows OTI results for the chemical midplane, $z_0$ (left), and mean vertical velocity, $v_{z0}$ (right), over the Galactic plane in cartesian coordinates of the star's guiding centre. The pale grey curve is the Line of Nodes found by \cite{Cabrera_Gadea_etal_2024} for Classical Cepheids. The black dot marks the Sun's position and the the radii $8,10,12,14~\text{kpc}$} 
    \label{fig:XY_g_OTI}
\end{figure*}

\section{Results}\label{s:results}

Figure~\ref{fig:XY_g_OTI} shows the midplane position, $z_0$ (left), and midplane vertical velocity, $v_{z0}$ (right), resulting from our OTI model fits for the sample of red giant stars in the low-$\alpha$ disc across bins of orbital radius and azimuth\footnote{Here, $(X_g,Y_g)=(R_g\cos\phi,R_g\sin\phi)$, and is computed for illustration purposes to show the results in a Cartesian projection proxy of guiding centre position.}. In addition, in the left panel we also overlay the line of nodes of the Milky Way disc warp determined using classical Cepheids from \citet{Cabrera_Gadea_etal_2024}. Focusing on the left panel of Figure~\ref{fig:XY_g_OTI}, we immediately see two things: 1) within an approximate radius of $R_g\lesssim10$ kpc, the value of the midplane position is $z_0\approx0$, indicating that within these inner regions the disc is flat and aligned with the reference frame; 2) beyond $R_g\gtrsim10$ kpc, the magnitude of $z_0$ becomes much greater ($z_0\approx0.2$), and the sign of $z_0$ changes predominantly with azimuth for constant orbital radius. In other words, at larger orbital radii there is a clear signature of a warp that changes primarily with azimuth, but also to some degree with orbital radius. The distribution of midplane vertical velocity measurements across orbital radius and azimuth (right panel of Figure~\ref{fig:XY_g_OTI}) reveals the expected picture for the Milky Way warp, i.e. positive vertical velocity over the line of nodes region. The value of $v_{z0}$ is approximately $0$ km s$^{-1}$ within $R_g\lesssim10$ kpc. At larger radii ($R_g\gtrsim10$ kpc), though, the magnitude of $v_{z0}$ increases strongly to $v_{z0}\approx10$ km s$^{-1}$. However, this is approximately the same across all azimuth bins. In summary, our OTI measurements reveal the warped distribution of the disc traced by RGB stars in the outer disc ($R_g>10$~kpc). Interestingly, the regions of the disc where our method infers $z_0\approx0$ kpc seem to line up well with the line of nodes determined from previous studies, attesting to this ``twisted'' pattern \citep{Chen3Dmap,Dehnen_2023_Warp_Ceph,Cabrera_Gadea_etal_2024,Poggio_great_wave_2024}. Although previous works have found indications of a twisted line of nodes in the general disc population \citep{RG19,McMillan_Wap_2024}, we argue that this work is the first to have recovered this feature empirically, thanks to the use of element abundance gradients.

\begin{figure*}
	\includegraphics[width=\textwidth]{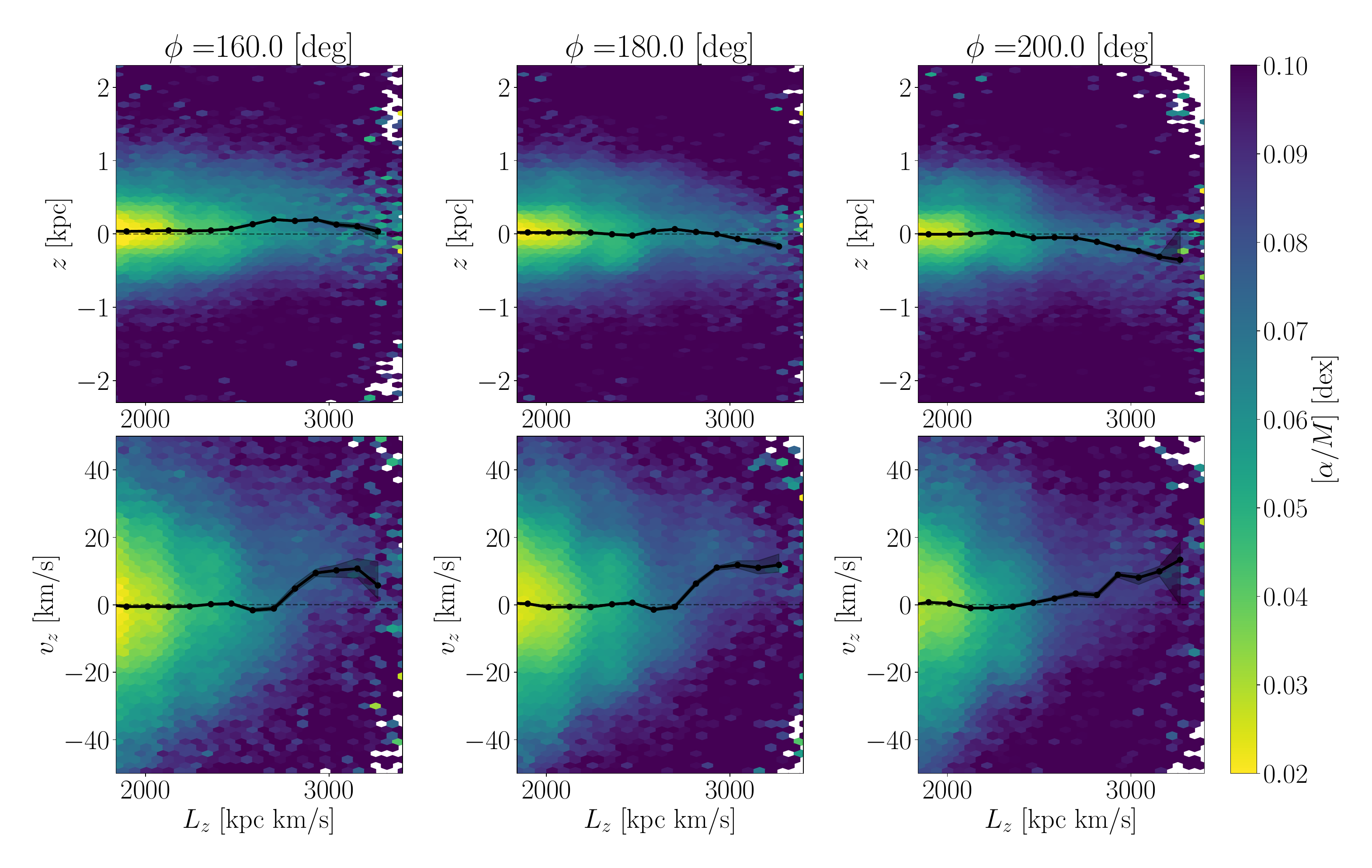}
    \caption{Median of the $\aom$ abundance (colour scale) over the $(z-L_z)$ (top panels) and $(v_z-L_z)$ planes (bottom panels) for three azimuthal bins centred on $\phi = 160^\circ$,$180^\circ$, and $200^\circ$ (columns from left to right). Only cells with more than 15 stars are shown. The black curve shows the OTI results. The shaded band spans the 16$^{th}$-84$^{th}$ percentiles from 50 bootstrapping realisations of the OTI results. The dashed horizontal line marks $Z = 0$ and $V_z = 0$ for reference.}
    \label{fig:Rg_ZVZ_alpha}
\end{figure*}

To investigate the distribution of $\alpha$-elements and its connection with the warp of the Galactic disc, we examine its behaviour in the $(L_z,z)$ and $(L_z,v_z)$ planes. Figure~\ref{fig:Rg_ZVZ_alpha} shows these distributions for three azimuthal bins centred at $160$, $180$, and $200$~deg, each $15$~deg wide, with the $(L_z-z)$ plane in the top row and the $(L_z-v_z)$ plane in the bottom row. In each panel, the background colour map represents the median $[\alpha/M]$ abundance in each hexbin, while the black curve traces the centre of the distribution recovered by OTI for a given $L_z$ bin (namely, the $z_0$ and $v_{z0}$ measurements in each azimuth and $L_z$ bin). The grey shaded region corresponds to the $16$th and $84$th percentiles obtained from OTI fits to 50 bootstrap realisations of the sample. Two main features emerge from these maps. First, the warped structure of the disc is apparent from simple visual inspection of the median $[\alpha/M]$ maps; this is expected given the ordered vertical kinematic and chemical structure of the low-$\alpha$ disc \citep[e.g.,][]{Danny_OTI_aR_2026}. Second, OTI robustly recovers the minima of the distribution, thereby quantifying the morphology of the warp traced by the RGB population. A similar behaviour is found for metallicity (Figure~\ref{fig:Rg_ZVZ_metal} in the Appendix B), although the signature is less pronounced owing to its weaker correlation with vertical dynamics, as previously reported by \citet{Danny_OTI_az_2024}. The results from OTI presented in Figure~\ref{fig:Rg_ZVZ_alpha} are reported in Table~\ref{table:results}, the mean $z_0$, $v_{z0}$ and their corresponding uncertainties.

\begin{table*}
\small
\caption{Results from each angular momentum bin at each azimuth.}  
\label{table:results}      
\centering                                    
\begin{tabular}{c c c c c c c c c c c c}          
\hline\hline                        
  $\phi$ & $L_z$ & $z_0$ & $v_{z0}$ & $P_{16}(z_0)$  & $P_{84}(z_0) $ & $P_{16}(v_{z0})$  & $P_{84}(v_{z0}) $   \\    

  [deg] & [kpc~km/s]& [kpc] & [km/s] & [kpc] & [kpc] & [km/s]& [km/s] \\
\hline                                 
    $160$&	$1700$ &	$0.031$ &	$-0.50$ &	$0.028$ & $0.034$ & $-0.86$ 	&$-0.17$  \\
    $160$&	$1800$ &	$0.04$ &	$0.12$ &	$0.039$ & $0.041$ & $-0.065$ 	&$0.38$  \\
    $160$&	$1900$ &	$0.034$ &	$-0.54$ &	$0.033$ & $0.036$ & $-0.78$ 	&$-0.29$  \\
\hline
\end{tabular}
\hfill\parbox[t]{\textwidth}{Description of the columns: $\phi$; Galactocentric azimuth. $L_z$; mean angular momentum. $z_0$; midplane position by OTI. $v_{z0}$; midplane vertical velocity by OTI. $P_{16}(z_0)$ and $P_{84}(z_0)$ are the $16^{\text{th}}$ and $84^{\text{th}}$ percentiles of $z_0$ from the bootstrap realisations. $P_{16}(v_{z0})$ and $P_{84}(v_{z0})$ are the $16^{\text{th}}$ and $84^{\text{th}}$ percentiles of $v_{z0}$ from the bootstrap realisations. The first few rows of the table are shown to provide guidance regarding its form and content. The full version of the table is available in the CDS.}
\end{table*}

The warp signal in $v_z$ as a function of $L_z$ recovered with the OTI method using $[\alpha/M]$ abundances is consistent with previous measurements obtained from different methodologies and stellar populations \citep{GEDR3_anticentre_2021,McMillan_2022_disturbed_outer_disc,Cheng_2020_Warp_patern}, reaching a maximum amplitude of approximately $10~\mathrm{km~s}^{-1}$. The corresponding signal in $z$ also follows the behaviour reported in the literature as a function of Galactocentric radius, notably crossing below the Galactic plane and displaying the characteristic arch-like morphology associated with the twist of the line of nodes \citep{Cabrera_Gadea_etal_2024}. The OTI determinations of $z_0$ and $v_{z0}$ are only weakly affected by low-number statistics. In the warped region ($L_z > 2500~\mathrm{kpc~km/s}$ or $R_g>11~\mathrm{kpc}$), the average separation between the 16th and 84th percentiles obtained from 50 bootstrap realisations is approximately $2~\mathrm{km/s}$ for $v_{z0}$ and $0.04~\mathrm{kpc}$ for $z_0$, indicating a high degree of robustness in the recovered warp signal.

\begin{figure*}
	\includegraphics[width=\textwidth]{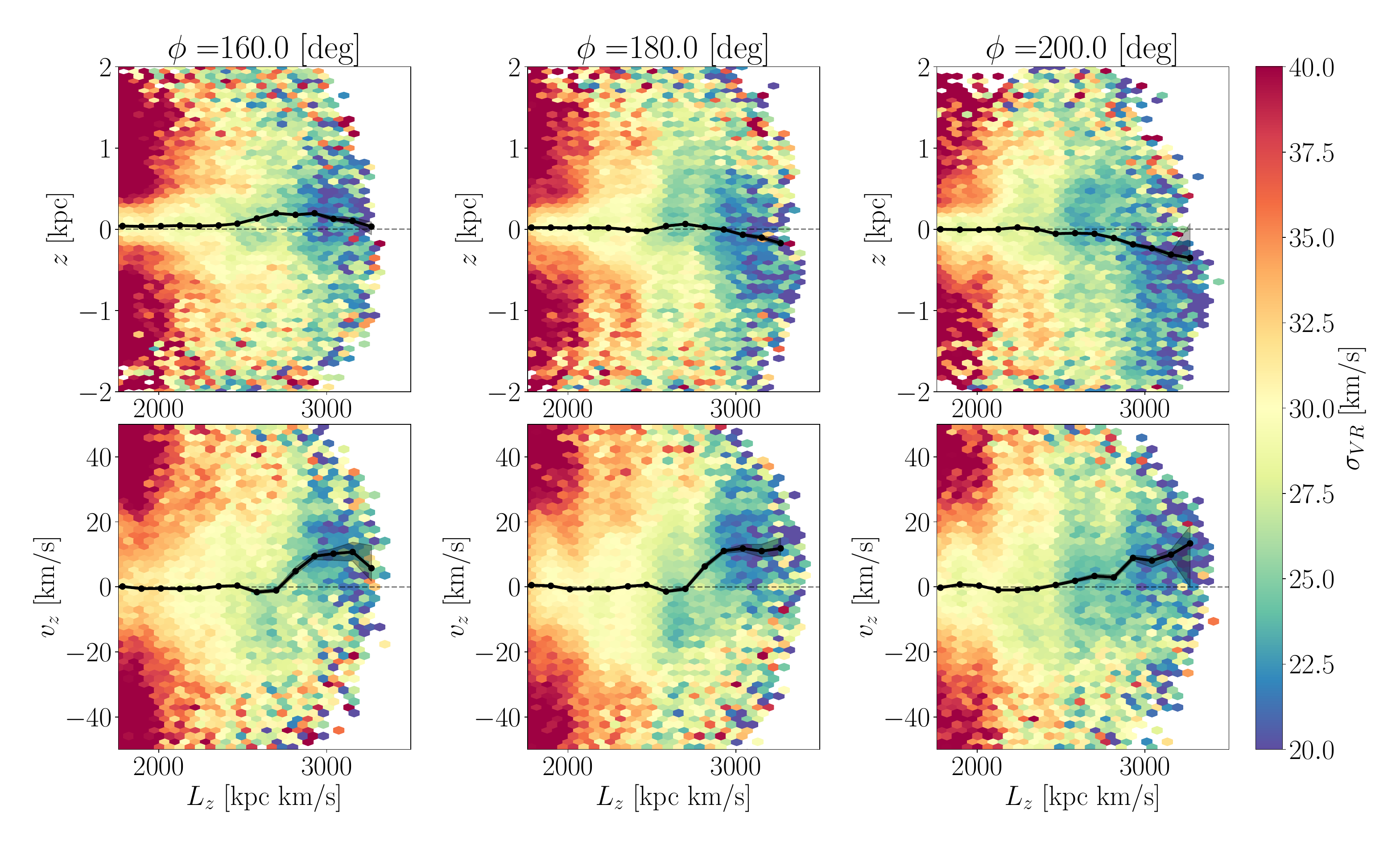}
    \caption{Same as Figure~\ref{fig:Rg_ZVZ_alpha} but in each bin the standard deviation of the Galactocentric radial velocity is shown. Only bins with more than 9 stars are shown.}
    \label{fig:Rg_ZVZ_VR}
\end{figure*}

To better characterise the kinematics of the stars in the warp, we explore the dispersion of $v_R$ in the planes $(L_z-z)$ and $(L_z-v_z)$, a quantity that, similarly to $\aom$, should also trace the Galactic warp. In Figure~\ref{fig:Rg_ZVZ_VR} we plot the same as in Figure~\ref{fig:Rg_ZVZ_alpha}, but in each cell of the planes we calculate the standard deviation of the radial velocity, $\sigma_{v_R}$. Although radial velocity dispersion is not as clear as the $\aom$ abundance, $\sigma_{v_R}$ shows a trend where regions with lower dispersion follow the warp signal (particularly clear for $v_z$) and therefore qualitatively agree with the OTI measurements. This last result is not surprising given that for the OTI measurements we have reduced the sample to those that have lower radial velocity ($|v_R|<50~\text{km/s}$). This result further emphasises that the warped region of the disc is populated by colder orbits.

\begin{figure*}
	\includegraphics[width=\textwidth]{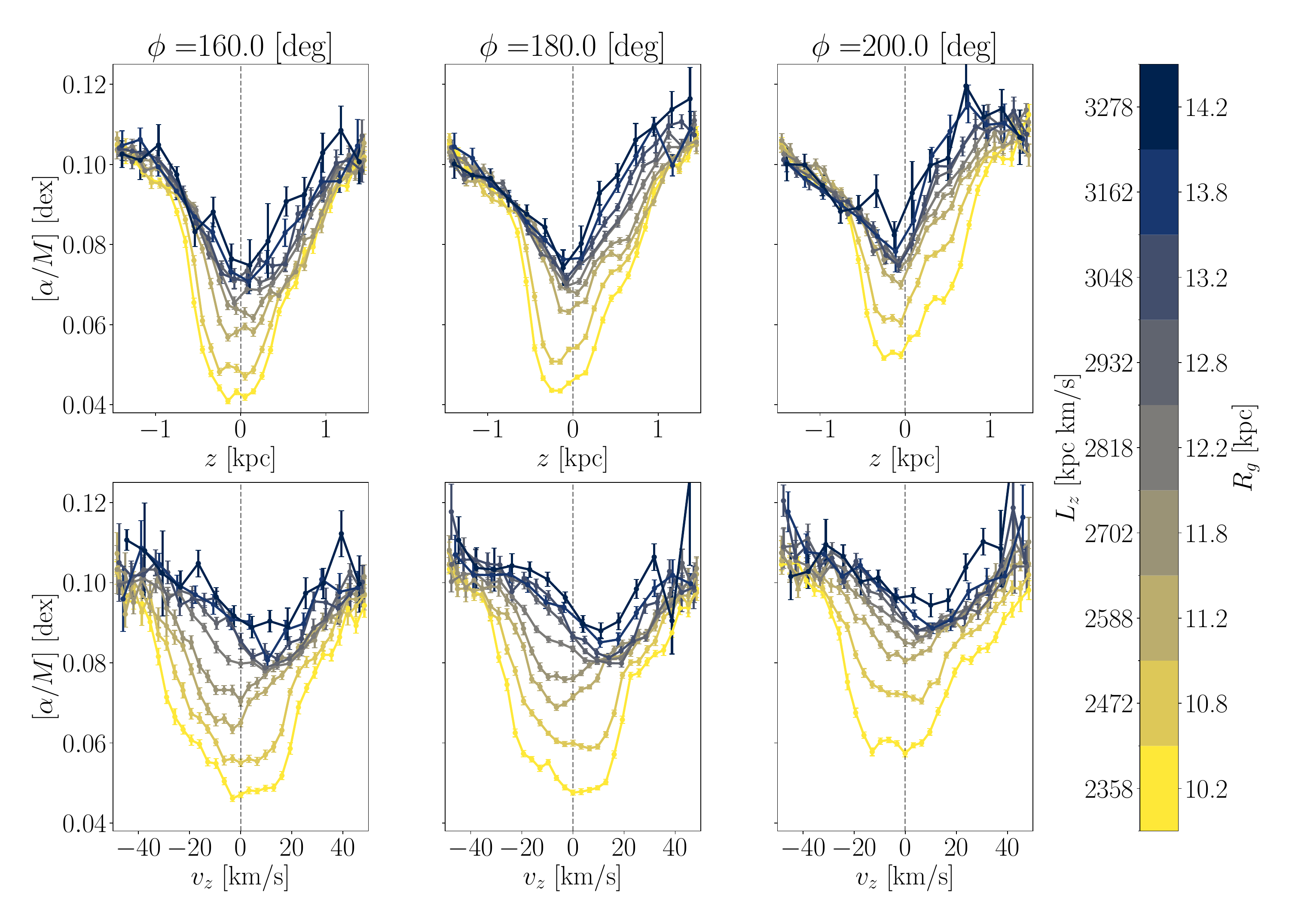}
    \caption{Each panel shows the median of $\aom$ abundances for each bin of $z$ (top panel) and $v_z$ (bottom panel) for different azimuths $\phi=160^\circ,180^\circ,200^\circ$ (per column, from left to right). The different curves in each panel show the mean for each $L_z$ bin (for convenience, we show to which guiding centre corresponds by taking $V_c=230$~\kms). The error bars in each result are the 16th-84th percentiles computed from $50$ bootstrapping realisations of the sample.}
    \label{fig:perfil_ZVZ_alpha}
\end{figure*}

In Figure~\ref{fig:perfil_ZVZ_alpha} we reproject our results to more visually illustrate the dependence of the saddle point in the $\aom$ abundances in the plane of vertical kinematics (either w.r.t. $z$ or $v_{z}$) with orbital radius, $R_g$ or $L_z$, for a given bin of azimuth. In this Figure, each column corresponds to a given azimuth bin, and each panel illustrates the median $\aom$ abundance profile, where it is apparent that the inflexion point in $\aom$ shifts depending on the orbital radius and vertical position or velocity. Instead of modelling the $(z-v_z)$ plane with OTI, we could have modelled the inflexion point of these median trends using a Gaussian or sech$^{2}$ model, as done in Tasso et al. in prep. Interestingly, the skewness or asymmetric distribution of alpha elements with respect to its minimum is evident up to $R_g\approx12$ kpc (or $L_z\approx2700~\text{kpc km/s}$). In these angular-momentum bins, the $\aom$ distribution even shows several inflexion points, which could be associated with the phase-space spiral location \citep{Antoja_2018_Phase_spiral}. In the outer disc, beyond $R_g\gtrsim12$ kpc (or $L_z\gtrsim2700~\text{kpc km/s}$), these median abundance asymmetries appear where $z_{0}>0$ or $v_{z0}>0$. These asymmetries (or lopsidedness) have a magnitude of approximately $0.01~\text{dex}$. 

\begin{figure*}
	\includegraphics[width=\textwidth]{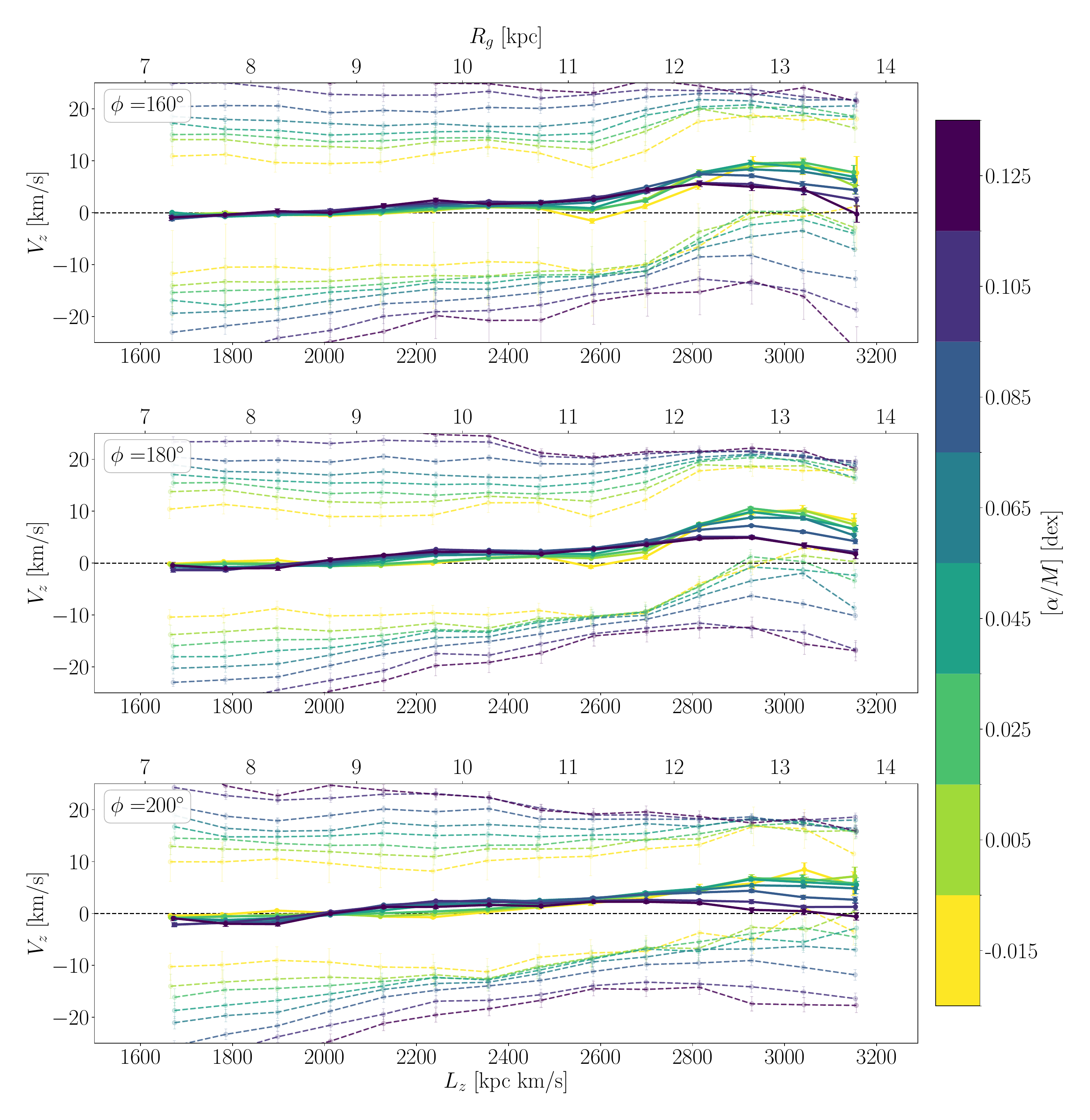}
    \caption{Mean vertical velocity as a function of angular momentum $L_z$ for mono-$\aom$ abundance populations for each azimuthal window (solid lines). The dotted lines for each mono-$\aom$ abundance populations show the $16$th and $84$th percentiles of the $v_z$ distribution. The error bars show the $16$th and $84$th of $50$ bootstrap realisations.}
    \label{fig:mono_abundance}
\end{figure*}

\begin{figure}
	\includegraphics[width=8.3cm]{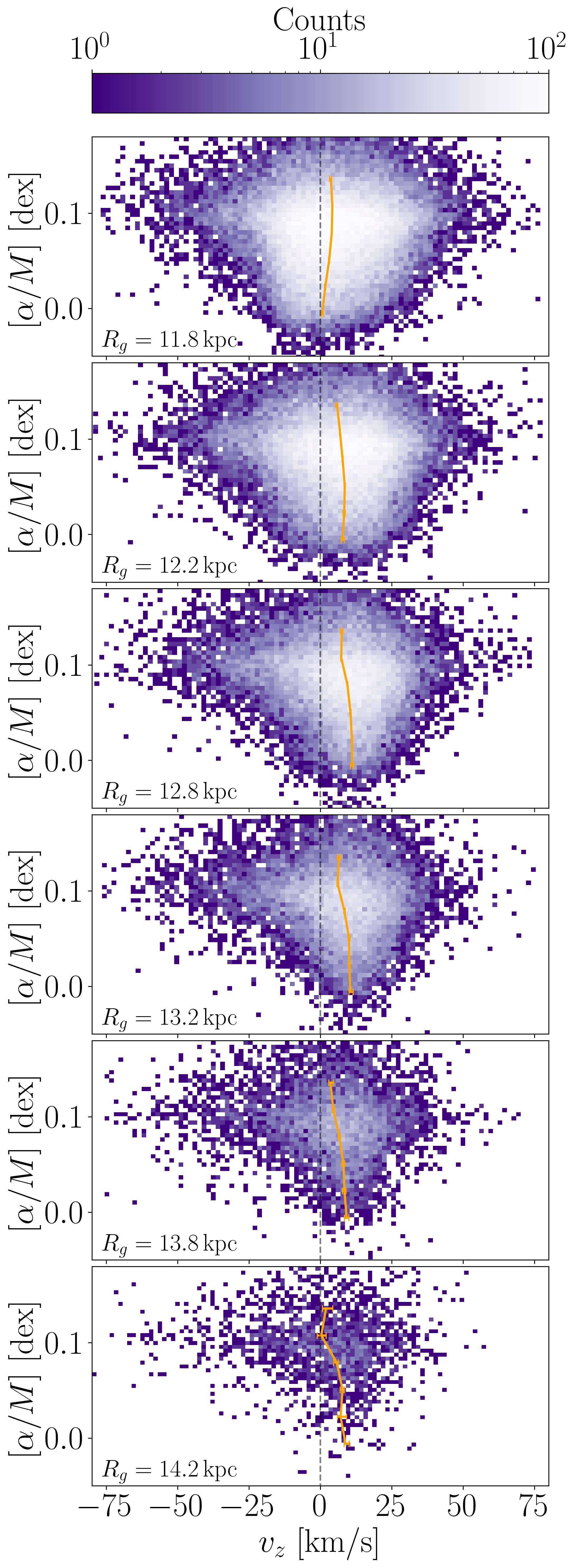}
    \caption{2D histogram over the $v_z$-$\aom$ plane for the anticentre sample in different angular momentum bins, where the associated orbital radius ($R_g=L_z/230\text{ km/s} $) is shown for reference. The solid orange line shows the mean $v_z$ for bins in $\aom$; the error bars are the $16\%$ and $84\%$ of $50$ bootstrapping realisations.}
    \label{fig:mono_abundance_dist}
\end{figure}

These asymmetries or lopsidedness in the $\aom$ distribution over the $(z-v_z)$ plane should give rise to different mean $z$ and $v_z$ for mono-$\aom$ populations, and therefore, for populations kinematically cold or hot. In Figure~\ref{fig:mono_abundance} we present the mean (solid lines), the $16$th and $84$th percentiles (dotted lines) of $v_z$ for different mono-$\aom$ populations as a function of angular momentum for our three azimuthal directions, the error bars are the $16$th and $84$th percentiles of each value presented from $50$ bootstrap realisations. We do not analyse the mean $z$ since it is more biased due to selection function effects and extinction by dust. The results from this examination show that the Galactic disc warp's signal in vertical velocity decreases for more enhanced $\aom$ populations in the outer disc; the signal also becomes more monotonic for $\aom$-rich populations. All curves show a growing amplitude up to a maximum around $R_g\approx 12.6~\text{kpc}$ (or $L_z\approx2900~\text{kpc km/s}$), where beyond this orbital radius the signal decreases. A similar behaviour has been found in the warp signal as a function of $R$ for different ages, with older populations showing a less pronounced warp signal \citep{Cheng_2020_Warp_patern} or for different populations chosen in the colour-magnitude diagram as a function of $L_z$ \citep{GEDR3_anticentre_2021}. For the distribution itself, it is interesting to notice how the $84$th percentiles for different populations get close together, and the opposite happens for the $16$th percentile in the warped regions, this differences in the distribution of each population give rise to the lopsided distribution of the full sample. This point is further remarked by Figure~\ref{fig:mono_abundance_dist}, where we show the 2D histogram over the $v_z,\aom$ plane for the anticentre sample ($\phi=180^\circ$) at different angular momentum bins (we report the orbital radius $R_g=L_z/230\text{ km/s} $). This Figure shows the $\alpha$-poor population distribution in $v_z$ is skewed towards negative velocities, making these populations the main factor in the asymmetries observed for the full population in Fig.~\ref{fig:perfil_ZVZ_alpha}.

\section{Discussion}\label{s:discussion}

\subsection{Element abundance gradients trace the Milky Way's warped disc}

To our knowledge, the results presented in this study are the first empirical (data-driven) measurements of the Galactic disc's warp using element abundance gradients (although see also Tasso et al in prep). Previous measurements of the Galactic disc's warp have predominantly centred on measuring the density of stellar tracers and computing their mean height with respect to the galactic plane (\citealp[e.g.,][]{Cheng_2020_Warp_patern,Skowron2019,Poggio_Warp_evolving,Uppal_2024}), particularly focusing on young stellar populations that are better tracers of the Milky Way's gaseous disc (\citealp[e.g., Cepheids:][]{Cheng_2020_Warp_patern,Skowron2019,Dehnen_2023_Warp_Ceph,Cabrera_Gadea_etal_2024}). There is one previous work that established a link between the chemistry of thin disc stars and the warp structure by analysing the metallicity of red clump stars in the \textsl{LAMOST, APOGEE}, and \textsl{Gaia} surveys \citep{Sun_2025_Fe_warp}. This study found that the height of the metallicity excess as a function of the radius in the Galactic disc (namely, $\Delta[\text{Fe}/\text{H}]$) resembles that traced by the mean height of Classical Cepheids. However, no direct measurement of the disc's midplane had ever been performed using element abundance gradients of disc stars.


The fact that element abundance gradients trace the orbital structure of the Milky Way disc, and that we can model this relationship with OTI, implies two important points about this Galactic component: $i$) the disc is highly structured in terms of its orbital evolution and chemical enrichment history, otherwise we would see no correlation between orbits and element abundances; $ii$) the level/magnitude of dynamical disequilibrium in the Milky Way disc is low, otherwise we would not be able to contour the ellipses of constant element abundance with OTI.

\subsection{Comparison with other warp tracers}

\begin{figure*}
	\includegraphics[width=\textwidth]{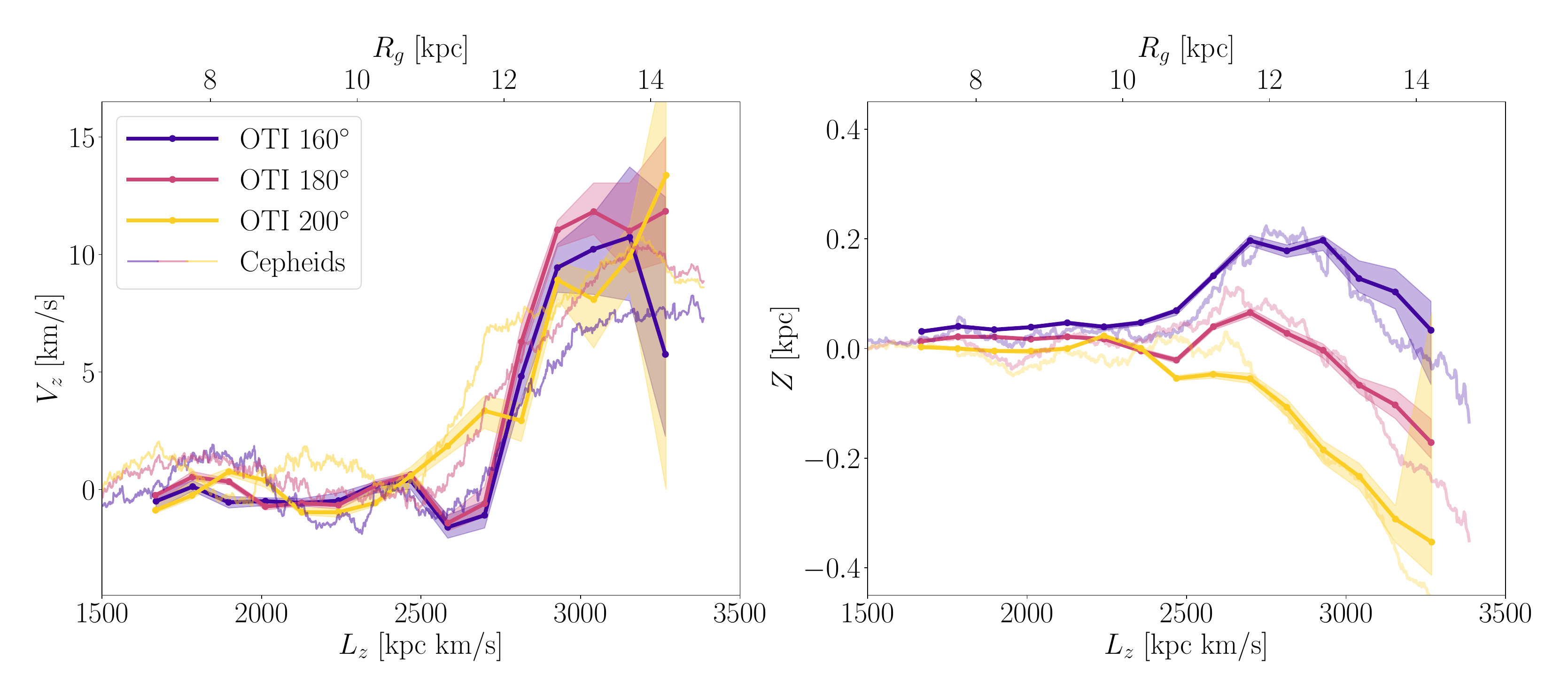}
    \caption{Mean vertical velocity and vertical chemical position determined by OTI (solid lines) for different Galactocentric azimuths as a function of angular momentum. The pale-coloured curves correspond to the mean vertical height of the Classical Cepheids reported by \cite{Cabrera_Gadea_etal_2024} at the same azimuths as the OTI measurements. The shaded coloured areas show the area between the $16\%$ and $84\%$ of 50 bootstrapping realisations.}
    \label{fig:Rg_Z}
\end{figure*}

The resulting $z_0$ and $v_{z0}$ measurements we determine for different bins of azimuth and orbital radii are summarised in Figure~\ref{fig:Rg_Z}, and are compared to results modelling the mean $z$ and $v_z$ of classical Cepheids from \cite{Cabrera_Gadea_etal_2024} by converting their radius to $L_z$ by means of the rotation curve from \citet[solving $L_z=RV_c(R)$]{Eilers_2019_rotation_curve}. Overall, there is excellent qualitative and quantitative agreement between our $z_0$ results modelling older RGB stars with OTI when compared to Classical Cepheids, emphasising that the centre found by OTI is close to that of the kinematically cold populations, such as Cepheids. Conversely, while the overall profile is similar, the qualitative resemblance between our OTI $v_{z0}$ measurements and those from Cepheids does not agree as well quantitatively. We postulate that a possible reason for this mismatch is that, at these azimuths, the kinematic signal of the warps is at its maximum (due to the line of nodes location \citealp{Chen3Dmap,Dehnen_2023_Warp_Ceph,Cabrera_Gadea_etal_2024}), therefore, any differences present between the signals are at their maximum expression in this region. However, there is broad qualitative agreement between our results and those from Cepheids.

An important topic in Milky Way warp characterisation is whether populations with different ages trace a different warp; various works have explored this issue and there is no consensus, as pointed out in \cite{Cabrera_RRLyrae_2025} (see Figure 8 in \citealp{Cabrera_Gadea_etal_2024} for a comparison of warp models with different tracers). These inconsistencies can arise from various factors, including distance determination, age determination, or how the warp is modelled and/or fitted. We explore this problem by taking mono-abundance populations, that at constant orbital radii should correlate with mono-age populations in the low-$\alpha$ (thin) disc \citep{Mackereth_age_metallicity_MW_2017, Bovy_Structure_Mono_abundance_2012}. 

As seen in Figure~\ref{fig:mono_abundance}, a smooth transition between mean $v_z$ for the different mono-$\aom$ populations occurs, the warp shows a clear decrease in $v_z$ amplitude as the population gets $\alpha$-rich, suggesting a decreasing amplitude for older stellar populations or, at least, kinematically hotter populations. This difference in the warp signal as a function of $L_z$ was already noticed by \cite{Cheng_2020_Warp_patern}, \cite{Wang_LAMOST_Disk_Warp_RC_2020} and \cite{GEDR3_anticentre_2021}; although they do not analyse the trend in chemistry but in age, or a proxy of it. Our results follow the same trend, particularly the results by \cite{Cheng_2020_Warp_patern} if we adopt that older populations correspond to those richer in $\alpha$ elements. We should remark the following distinction: previous works on the change in amplitude of the warp for different stellar populations \citep{Evolution_disc_shape_Amores_2017,RG19,Uppal_2024,Cheng_2020_Warp_patern,Chrobacova_GDR2_Warp_Model_2020GDR2,Wang_LAMOST_Disk_Warp_RC_2020,Li_2023_Nonsteady_Warp,Cabrera_RRLyrae_2025} mainly analysed the mean $z$ or $v_z$ as a function of radius; therefore, the differences between those signals for old and young populations might arise as a consequence of the epicycle of the stars blurring the original signal imprinted in $L_z$, which should be cleaner and more similar between populations if analyses as a function of $L_z$. The persistent difference and smooth gradient between stellar populations in the warp signal as a function of $L_z$ suggest that populations reacted or evolved differently to the same gravitational perturbations, and the differences between them have reasons beyond the blurring due to the epicycle; it might also suggest that the coupling between in-plane and vertical motion plays a role in how the signal was imprinted and evolved.  

\subsection{The asymmetric distribution of $\aom$ elements in the vertical phase space}

In Figure~\ref{fig:perfil_ZVZ_alpha} we presented how the profile of the median $\aom$ abundances changes as a function of $z$ and $v_z$ for different bins of angular momentum and azimuth, and how these profiles are asymmetric with respect to their inflexion point (or minimum $\aom$). For the inner disc, we argue that these features are expected as a consequence of dynamical disequilibrium features such as the phase-space spiral, since it is also present in the chemical distribution \citep{Frankel_Iron_Snail_2025}. It is also possible that the changes in the asymmetric profiles of average $\aom$ with azimuth, at constant $R_g$ observed in Figure~\ref{fig:perfil_ZVZ_alpha}, are features expected from the phase spiral rotations observed by \cite{Alinder_2023_rotation_phase_spiral} and \cite{ Widmark_2026_PhaseSpiral_rotation}. Conversely, for the outer disc beyond $L_z\approx2600~\text{kpc km/s}$, the signal of the phase spiral is hardly recognisable in the \textsl{Gaia} data \citep{Antoja_PSinDR3_2023,Alinder_2023_rotation_phase_spiral}. Thus, while these asymmetries with respect to the inflexion point of the median $\aom$ profile are still present, these could arise because we are looking at the beginning of phase mixing. Since dynamical times are longer at those larger radii, it is possible that the phase-space spiral has not yet formed (or wound enough) in those regions and it is only starting to form, causing the lopsidedness we see in Figure~\ref{fig:perfil_ZVZ_alpha}. 

These asymmetries in the median $\aom$ must have their counterpart in $f$ or, in particular, the vertical density profile. The work by \cite{Wu_2026_asymetric_density_porfile} analysed the vertical density profile of the (thin) disc as a function of radius and azimuth traced by red clump stars by fitting a exponential profile to \textit{Gaia} DR3 data, taking into account the selection function. The authors found that the density peak shows a similar trend as the warp, and that the exponential profile does not capture the full complexity of the vertical distribution in the disc outskirts ($R>11~\text{kpc}$), an asymmetric distribution with respect to the density peak can be observed. We argue that this result by \cite{Wu_2026_asymetric_density_porfile} corroborates our findings and suggests that the disc is, at first order, bent rigidly in its outskirts, keeping most of its $(z-v_z)$ plane structure even within the warped region of the stellar component. This view is further supported by the fact that the colder orbits are kept within the warped region as shown by Fig~\ref{fig:Rg_ZVZ_VR}. {Some of these features have been observed in warp simulations. \cite{No_memory_warps_2023} reported that, in their cosmological simulations, warps formed while largely preserving their vertical density distribution. However, differences were found between warps composed mainly of accreted stars and those formed from pre-existing disc stars. The former showed larger deviations from a double-$\mathrm{sech}^2$ vertical density profile, whereas in the latter the discs bent more rigidly.}

In connection with the differences in the warp signal for different mono-$\aom$ populations, we would like to offer a plausible explanation  for how they might be connected, albeit somewhat speculative. If we have in mind the analytical explanation by \cite{Binney_phase_spiral_2018} of the phase spiral, once the disc is moved out of equilibrium, each population will land on a different range of vertical frequencies; since they extend differently on the $(z-v_z)$ plane, colder populations will land on a smaller frequency range than hotter populations. As a consequence of this, different mono-$\aom$ populations will mix differently, and therefore, as their distribution evolves in the $(z-v_z)$ plane, they will develop a different mean $z$ and $v_z$. Another side of the explanation of the lopsidedness of alpha elements in the $(z-v_z)$ plane is that in a vertical perturbation to the stars, some orbits will be more resonant to the perturbation than others, and also, the in-plane motion, and radial velocity dispersion plays a role on the signal of the warp imprinted \citep{Sellwood_disk_warps_2013,Khachaturyants_gas_inflow_warp_2022}, in particular the warp signal is richer in structure for cold populations since they can support shorter wavelength bending modes \citep{Toomre1966,Khachaturyants_gas_inflow_warp_2022}. We will explore these hypotheses in future work by means of toy models and N-body simulations.

\section{Conclusions}\label{s:conclusion}

In this work, we have explored the Milky Way warp by analysing the $\aom$ distribution of RGB stars in the \textsl{Gaia} DR3 data from the \cite{Li_AspGap_RGB_2024} catalogue. Our main conclusions are:

\begin{itemize}
    \item The $\aom$ element abundance distribution can be used to trace the Milky Way's disc warp; the OTI method together with $\aom$ abundances delivered using \textsl{Gaia} XP data is capable of tracing the warp up to $L_z\approx 3300~\text{kpc km/s}$ or $R_g\approx 14~\text{kpc}$. The warped shape of the distribution follows closely that of Classical Cepheids in $z$, even revealing the twisted line of nodes, although in $v_z$ some quantitative differences are present.
    \item An asymmetric distribution of $\aom$ abundances with respect to the inflexion point in $z$ and $v_z$ is found. 
    This asymmetry changes as a function of $L_z$ and azimuth. For the inner disc, this asymmetric feature could be attributed to the phase spiral; for the outer disc, such asymmetries might be due to the beginning of phase mixing in the disc in vertical kinematics (i.e., the arising of the phase-spiral).
    \item Mono-$\aom$ abundance populations show differences in their mean $v_z$ as a function of angular momentum $L_z$. For richer $\aom$ abundance populations (kinematically hotter populations), the $v_z$ signal as a function of $L_z$ decreases and becomes more monotonic. This result leads to conclude that the differences in the warp signal between different stellar populations as a function of $R$ is more than a signal blurred by the epicyclic excursions.
\end{itemize}

The exploration of the Milky Way disc's warp with chemistry opens a new view into its structure and kinematics, but also provides an independent and less bias prone avenue to analyse its dynamics. The fact that within the transition zone between the flat and the warped disc, and also within the warped regions, the disc keeps its chemical gradients in the $(z-v_z)$ plane is a new feature to explore in simulations for the different mechanisms that could form the Milky Way warp. Furthermore, current/future large-scale spectroscopic campaigns (e.g., \textsl{SDSS-V/Milky Way Mapper}: \citealp{Kollmeier2026}; \textsl{WEAVE}: \citealp[]{Jin_2023}, amongst others including future \textsl{Gaia} data releases) will supply the necessary data and coverage to repeat these measurements across a broader area of the Milky Way disc revealing more details of the warp.

\section{Data availability}

Table \ref{table:results} is only available in electronic form at the CDS via anonymous ftp to cdsarc.u-strasbg.fr () or via -- .

\begin{acknowledgements}
This research has been supported by funding from project FCE\_1\_2021\_1\_167524 of the Fondo Clemente Estable, Agencia Nacional de Innovaci\'on e Investigaci\'on (ANII). This work has made use of data from the European Space Agency (ESA) mission {\it Gaia} (\url{https://www.cosmos.esa.int/gaia}), processed by the {\it Gaia} Data Processing and Analysis Consortium (DPAC, \url{https://www.cosmos.esa.int/web/gaia/dpac/consortium}). Funding for the DPAC has been provided by national institutions, in particular the institutions participating in the {\it Gaia} Multilateral Agreement. DH is supported by the UKRI Science and Technology Facilities Council under project 101148371 as a Marie Skłodowska-Curie Research Fellowship. \\  

{\it Software:} Astropy \citep{astropy2018},
    AstroML \citep{astroML},
    Matplotlib \citep{mpl},
    Numpy \citep{numpy},
    Jupyter \citep{jupyter2016}, 
    TOPCAT \citep{Topcat2005,Stilts2006} and
    PYGAIA (\url{https://github.com/agabrown/PyGaia}).

\end{acknowledgements}

\bibliographystyle{aa}
\bibliography{warp_refs}


\begin{appendix}

\section{A. Effects in $\aom$ distribution due to selection function}\label{Ap:A}

In this section we explore possible effects of the selection function in our results. 

Since the metallicity of RGB stars correlates with luminosity (decreasing metallicity increases brightness), and there is a correlation between metallicity and $\aom$ abundance in the thin disc, this will introduce a correlation between luminosity and $\aom$ abundance. As the Gaia selection function imposes a fixed apparent brightness, combined with the effect of extinction, this may affect the range of chemical abundances present in our sample for different azimuths at a given radius.

\begin{figure}
	\includegraphics[width=9cm]{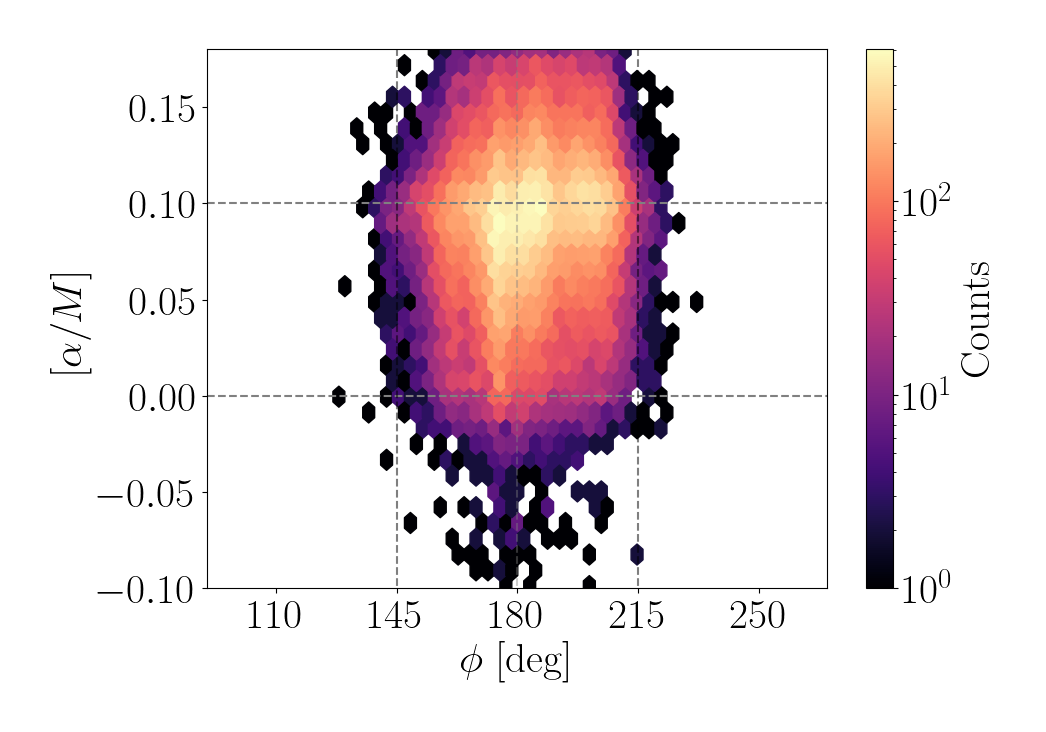}
    \caption{2D histogram of the RGB sample in the plane $\phi,\aom$. The black dotted vertical lines show the azimuth range of our bins. The horizontal lines show the cuts in $\aom$ for the restricted sample.}
    \label{fig:alpha_Phi}
\end{figure}

In order to characterise whether the observed warped chemical distribution is an artefact of these correlations, we explore if the vertical gradients are still observed when restricting our sample to $0<\aom<0.1$. This cut ensures our sample maps the same range of abundances for the different azimuths without over representing any populations, and also guarantees that we map the bright stars. In Figure~\ref{fig:alpha_Phi} we show the 2D histogram of Galactocentric azimuth $\phi$ and $\aom$ for the sample in a radial bin $12.5<R/\text{kpc}<13.5$ (a similar trend is observed for other radial bins). The black dotted lines show the azimuth limits of our bins, and the grey dotted line shows $\phi=180^\circ$ for reference. In the range $\aom<0$, the V shape of the distribution centred at $\phi=180^\circ$ is a consequence of the change in heliocentric distance bias to detect the $\aom$ poorest in the nearest zone because they are the less luminous. The figure illustrates that this bias affects only a very minor fraction of our sample, which spans quite a uniform range of $\aom$ in the azimuth range explored. We now further limit our sample to $0<\aom<0.1$ to ensure the same range of $\aom$ across azimuth and repeat the procedure used for the full sample as described in Sec.~\ref{s:method}.

\begin{figure*}
    \centering
	\includegraphics[width=0.9\textwidth]{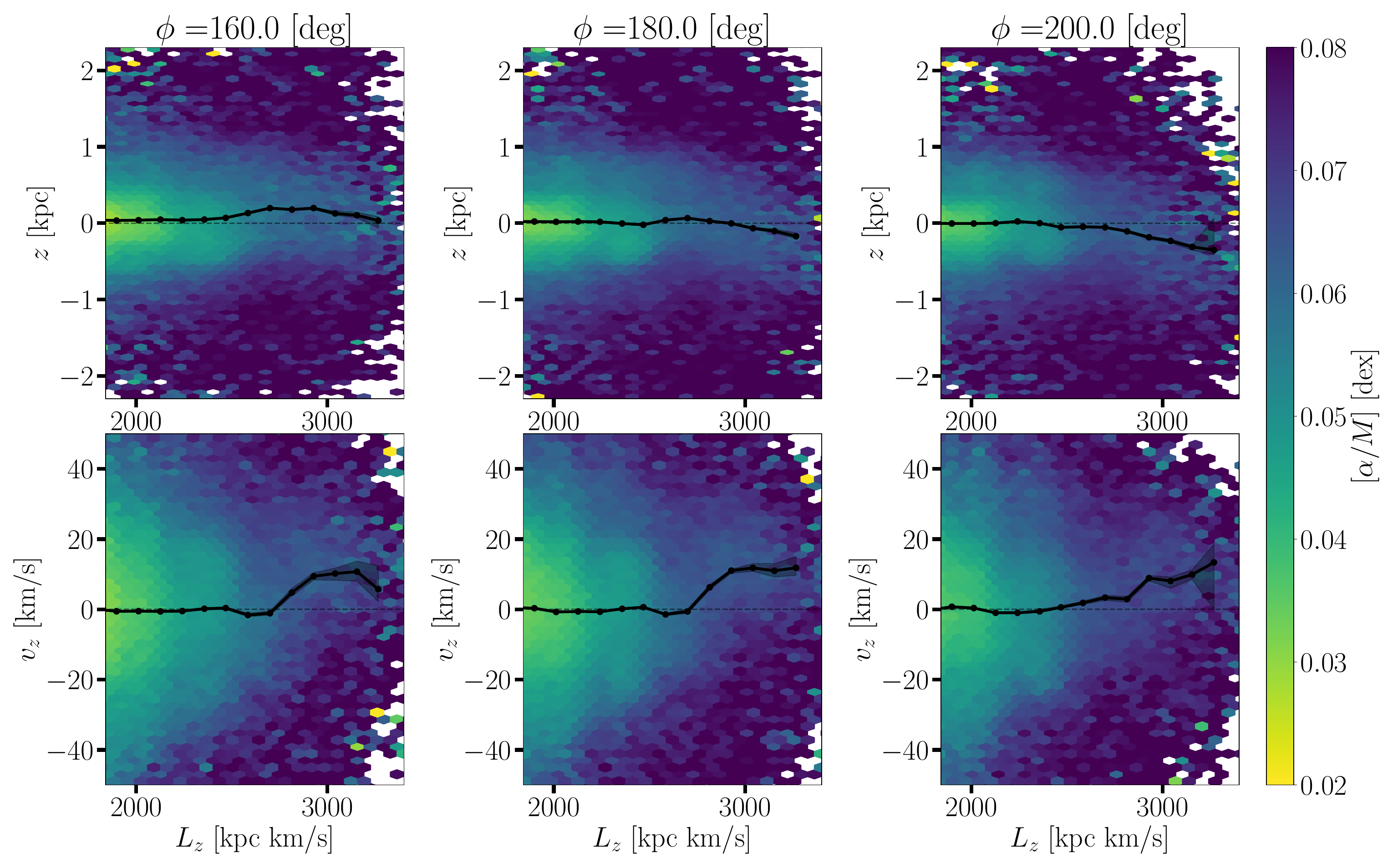}
    \caption{Median of the $\aom$ abundance for the limited sample $0<\aom<0.1$ over the planes $z,L_z$ (top panels) and $v_z,L_z$ (bottom panels) for three azimuthal bins centred on $\phi = 160^\circ$,$180^\circ$, and $200^\circ$ (columns). Only cells with more than 15 stars are shown. The black curve shows the OTI results for the full sample. The shaded band spans the 16th-84th percentiles from 50 bootstrapping realisations of the OTI results. The dashed horizontal line marks $z = 0$ and $v_z = 0$ for reference.}
    \label{fig:Rg_ZVZ_SF}
\end{figure*}

In Figure~\ref{fig:Rg_ZVZ_SF} we show the same planes as in Figure~\ref{fig:Rg_ZVZ_alpha} for the limited sample, the OTI results for the full sample are shown by the black lines. The figure shows that the warped distribution of the chemistry is present in both $z$ and $v_z$ with the same trend as the OTI results. These results confirm that the warped distributions of $\aom$ and $\moh$ are not an artefact due to selection function effects.

\section{B. The warped distribution of $\moh$.}

\begin{figure*}
    \centering
	\includegraphics[width=0.9\textwidth]{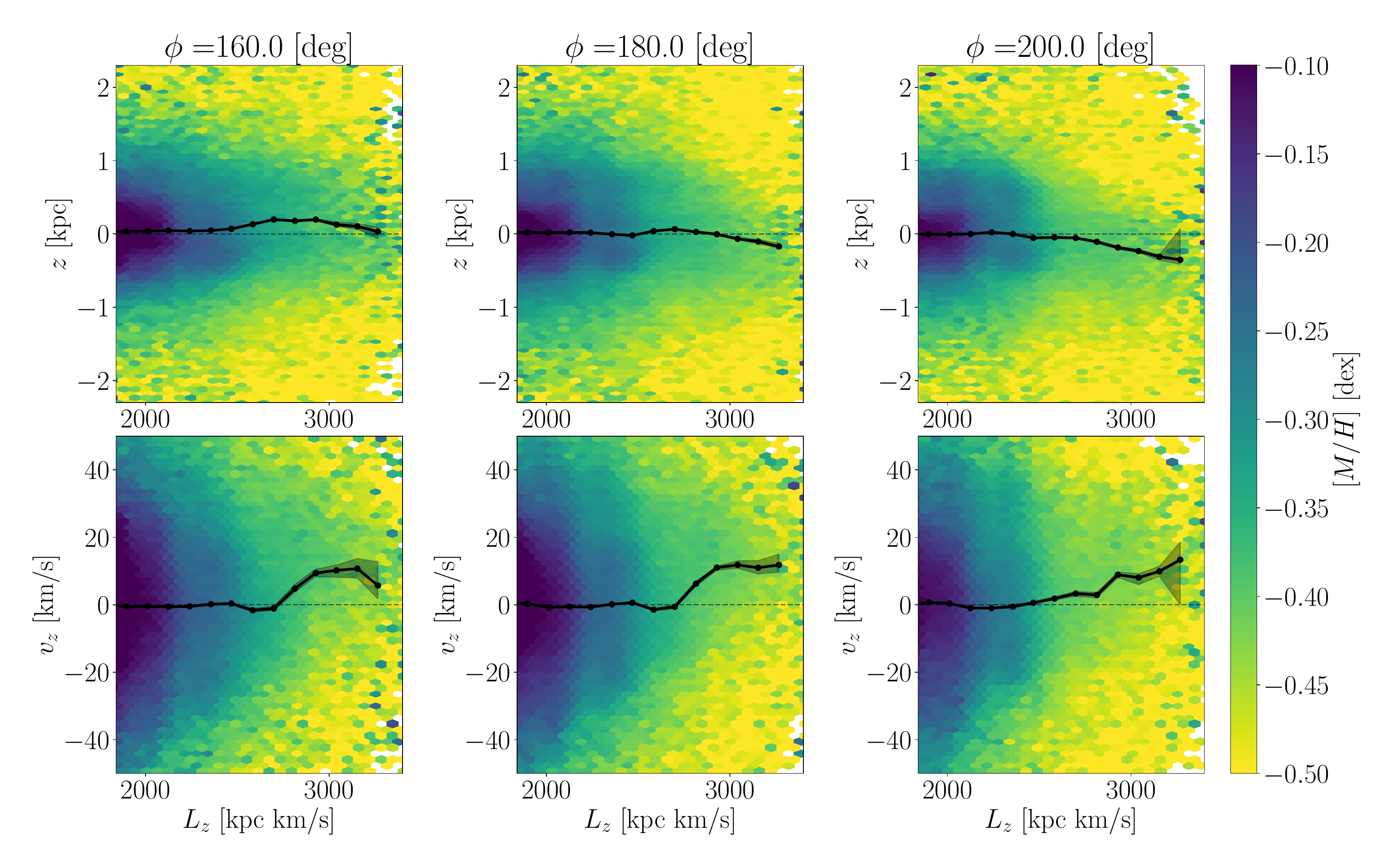}
    \caption{Median of the metallicity abundance over the planes $z,L_z$ (top panels) and $v_z,L_z$ (bottom panels) for three azimuthal bins centred on $\phi = 160^\circ$,$180^\circ$, and $200^\circ$ (columns). Only cells with more than 15 stars are shown. The black curve shows the OTI results. The shaded band spans the 16th-84th percentiles from 50 bootstrapping realisations of the OTI results. The dashed horizontal line marks $z = 0$ and $v_z = 0$ for reference.}
    \label{fig:Rg_ZVZ_metal}
\end{figure*}

In this section we show the distribution of metallicity in the  $L_z,z$ and $L_z,v_z$ planes in Figure~\ref{fig:Rg_ZVZ_metal}, similarly to Figure~\ref{fig:Rg_ZVZ_alpha} for the $\aom$. The black curve are the OTI result for the $\aom$ abundances. The metallicity distribution shows the warped distribution of the disc in $z$ and $v_z$, mostly dominated by a rich population as expected from the correlation between $\aom$ and $\moh$. The results by \cite{Sun_2025_Fe_warp} are corroborated by this result in our sample.

\end{appendix}

\end{document}